# Impact of Electron-Withdrawing Groups on Ion Transport and Structure in Lithium Borate Ionic Liquids

Volodymyr Koverga, Selvaraj Selva Chandrasekaran and Anh T. Ngo[*]

Department of Chemical Engineering, University of Illinois Chicago, Chicago, IL 60608, United States
Materials Science Division, Argonne National Laboratory, Lemont, IL 60439, United States

[*]Corresponding author's email: anhngo@uic.edu

**Abstract**

Among the distinctive structural features of lithium ionic liquids (LILs), a novel class of single-component electrolytes, the variation of the electron-withdrawing group stands out as a key factor in determining their dynamics. To understand this phenomenon, we conducted molecular dynamics (MD) simulations for LILs based on hexafluoro-2-propanoxy (LIL2), hexafluoro-2-methyl-2-propanoxy (LIL4), and trifluoro-2-propanoxy (LIL6) derivatives. Results revealed that correlated ion dynamics govern the general transport characteristics in LILs, while the electron-withdrawing group regulates the Li transport mechanism. Upon saturation by fluorine atoms, LILs exhibit higher inhomogeneity in their transport and structure properties. Strong coordination along the ethoxide group promotes jumps of Li across positive domains, while in fluorine-poor LILs, stronger coordination in proximity to boron atoms carries the anion along Li transport. Understanding the results of MD simulation will aid the further design and widespread use of this class of electrolytes in production of the energy storage and conversion devices

**TOC Graphic**

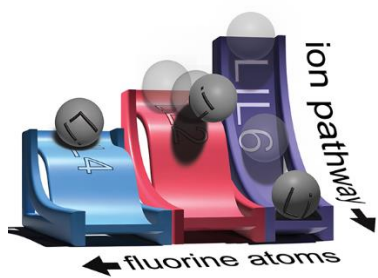

Since the first commercial Li-ion batteries (LIBs) hit the market in the early 1990s, their development has transcended the realm of portable electronics. Their inherent features, including high specific capacity, voltage, lack of memory effect, excellent cycling performance, minimal self-discharge, and wide temperature range of operation,[1, 2] make them superior to other energy storage systems in the battery market today. Nevertheless, as material advancements approach their limits, the escalating demand for more cost-effective and high-performance devices requires a more rational approach in the development of the new generation of LIBs.[3]

In this context, electrolyte design is a fundamental challenge in battery electrochemistry.[4] Apart from the well-known requirements of electrolytes for conventional LIBs, such as high conductivity, chemical and electrochemical stability, low flammability, and environmental sustainability, the electrolytes must also prevent the formation of lithium dendrites by forming a passivating solid electrolyte interphase towards the metal anode.[5] Most current commercial LIBs use liquid electrolytes based on volatile and flammable organic carbonate solvents, which can easily undergo drastic degradation processes.[6] Moreover, as a result of the ion-solvent imbalance in these electrolytes the formation of concentration gradients interferes the LIB performance. Since high solvation of Li in carbonate electrolytes leads to relatively low Li transference number (typically 0.35-0.40)[7] and high negative space charge due to the anion motion. Therefore, for the increase of the Li transference number, ideally up to unit,[8] it is necessary to make a step forward in the development of solvent free/single-ion electrolytes.

To advance the research in this direction it was explored various approaches, including the utilization of electrolytes based on the metal-organic frameworks, polymers/polyanions, highly concentrated solutions or deep eutectic solvents. ILs has been found to be the most promising candidates for this role due to their non-flammability, low vapor pressure, high thermal and electrochemical stability, and containing the charge-balanced anions and cations.[9] Incorporating ILs as an electrolyte media in LIBs, however, hindrances the Li mobility due to the naturally high viscosity of ILs coupled with the presence of larger cation in solution.[10] To reduce the high viscosity, ILs are commonly admixed with organic electrolytes (or their mixtures), making the electrolyte more conducive via the solvation of Li and the dissociation of salts. As a result, the ion conductivity is increased from $10^{-3}$ to $10^{-2}$ S cm$^{-1}$,[11] although the Li-salt concentration polarization persists in the electrolyte due to the low transference number.

To completely suppress concentration polarization, solvent-isolated Li-salts, which (like common ILs) are liquids at room temperature, can be introduced as an alternative electrolyte in LIBs. Only a few such LILs have been reported to exist today, based on Li-salts of aluminates and borates, and Li coordinating ether ligands with electron-withdrawing groups. The first generation of Li$^+$Al(OR)$_4^-$ LILs was introduced by Fujinami and Buzoujima[12], where R represents two oligoether groups and $CF_3CO_2$, $CF_3SO_3$, $(CF_3SO_2)_2N$ or $C_6F_5$ were chosen as electron-withdrawing groups. They found that enhancement of the ionic conductivity (up to $10^{-4}$ S cm$^{-1}$ at 40 °C) in these electrolytes is achieved by the degree of salt dissociation, which is ascribed to the electron withdrawing ability of the groups anchored onto anion center. Moreover, a relatively high transference number of these LILs is mainly associated with incorporation of short-length oligoether chains inducing the anion immobilization. In attempts to ion mobility, Watanabe *et al.*[13, 14] reported another type of borate-based LILs, where $CH(CF_3)_2$, $C_6F_5$ and $CF_3CO$ are used as electron-withdrawing groups along with Li coordinating oligoether groups, so-called LiHFIP, LiPFP and LiTFA, respectively. Based on the detailed analysis of physicochemical properties of designed LILs, the authors noted a high viscosity as a main factor that limits ion conduction and increases the glass transition temperature. In the recent work of Shigenobu *et al.*[15] it is reported that the borate-based LILs, such as LiHFIP and LiFTA, can induce concentration polarization, which results in reduction of transference number. Using MD simulation approach Farhadian and Malek[16] tried to understand the nature of the low ion mobility in LiFTA. According to the reported results ion migration in the system can be described in terms of a hopping process, where the low ionic conductivity is explained by the low rate of cation migration from one cage to another and the long lifetime of ion pairs.[17]

Other works devoted to functionalization of Li$^+$B(OR)$_4^-$ paved the route for a further understanding of the chemistry of LILs and the crucial factors for their improvement as a potential electrolyte. Particularly, Zygadło-Monikowska *et al.*[18] obtained a new structure by substituting one of the oligo(ethylene oxide) by butyl group. It was reported that the ionic conductivity of the resulting salts depends on the number of ethylene oxide units, where the highest conductivity of $2 \times 10^{-5}$ S cm$^{-1}$ is exhibited by the salt with the chain containing three monomers. To gain insight into how the functionalization of LILs affects the ionic conductivity, Guzmán-González et al.[19] studied a design concept of a series of borate-based LILs with various electron-withdrawing groups and alkane substituents. These LILs showed high ionic conductivity values (higher than $10^{-4}$ S cm$^{-1}$ 25 °C) and transference numbers of 0.4-0.5, as well as a high compatibility with lithium-metal electrodes with stable polarization profiles. The authors emphasized the role of the balance between the electron-withdrawing capacity of the fluorinated groups, the solvating capacity of the ethoxide groups, and the interfacial compatibility of the stabilizing aliphatic groups that allow to achieve higher conductivity values.

Despite the obvious inexhaustible potential of the LIL systems mentioned above, to date, there is a limited number of works describing them by using atomistic level details. Therefore, in this work, we present the results of a computational insight into the effect of electron-withdrawing groups on the dynamics and local structure organization in the recently reported boron-based reported LILs.[19] Apart from the availability of experimental data on the ion dynamics for these LILs, we believe that the theoretical characterization of these systems in terms of their underlying Li transport, as well as understanding their structural features will have an impact for the molecular design of novel LILs with even higher mobility characteristics, which will be useful for future LIB development.

For this purpose, classical MD simulation was conducted for the set of LILs, originally introduced as LIL2, LIL4 and LIL6, which exhibited the highest ionic conductivity and transference number. Potential models for these systems were mainly adopted from OPLS-AA library[20, 21] and complemented by relaxed potential energy surface scans. Utilizing previous experience,[22-25] initially low ion dynamics were enhanced by tuning the van de Waals potentials for Li and H atoms to reduce the energetics between counterions and anions, respectively. Additionally, electrostatic potentials based on isolated ion pair calculations were further tested for their ability to increase the rate of ion dynamics using other popular approximations:[26-28] unit charges[29] and scaled by a factor of 0.8[30-32] calculated for isolated anion. All the results and remaining technical details are available in the Supporting Information.

Hence, ion conductivity was estimated using the Nernst-Einstein approximation and its expression with Onsager transport coefficients (further below "Onsager characteristics" for simplicity).[33] This approach not only enhances the accuracy of the calculated properties but also helps to understand the impact of individual, *self*, or collective, *correlated*, ion contributions on transport properties. An additional challenge in computing Onsager characteristics, particularly for collective contributions, is achieving the diffusive regime, which requires overcoming rather poor statistics of long-time displacements. The use of short-time statistics to estimate collective displacements is therefore often recommended, however, we refer to the fundamental work of Kubisiak and Eilmes, which thoroughly discusses the pros and cons of statistical processing of Onsager conductivity.[34] Therefore, the present study involved the correlation depth of 20% for all frames along the trajectory. In addition, the data sampling was improved by averaging over nine additional independent calculations. Therefore, the length of the diffusion regime along the trajectory was 10-15% for correlated displacement and 70-80% for individual displacements (Figure 1a-c).

Quantitative assessment of ion mobility characteristics against experimental measurements is displayed in Figure 1d-f. Results indicate that calculated ionic conductivities (Figure 1d), from two approximations, align with the experimental trend: LIL6 > LIL2 > LIL4. Specifically, while there are minor statistical deviations of around 5%, the ionic conductivities derived from considering individual ion contributions alone are significantly higher, about an order of magnitude, than in the experiment. Moreover, they are two orders of magnitude greater than conductivities involving collective ion contributions. In contrast, Onsager conductivities shows a relatively large deviations from the average ~30-35%, while discrepancies with experimental conductivity range ~36-48% for LIL6, LIL2 and LIL4. This difference between the two approximations highlights the importance of considering collective ion motion in determining ionic conductivity, which it can be quantified using the Haven ratio.[33] This quantity determines the deviation of collective ion contribution with respect to individual contribution or, in other words, Haven ratio expresses the degree of collective ion contribution to the ionic conductivity. The general trend for three LILs is the same as the one shown for ionic conductivity: LIL6 (0.09) > LIL2 (0.07) > LIL4 (0.05), where decreasing Haven's ratio suggests a trend towards less correlated ion motion and lower coordination between ions.

To better understand this phenomenon, we can directly trace each ion contribution to ionic conductivity. For simplicity, Onsager conductivity is decomposed and converted into diffusion coefficients of individual and collective terms. Figure 1e shows that individual cation mobilities in LILs are generally slightly higher by ~3-4% compared to corresponding anions. However, collective diffusion presents an opposite trend. Indeed, the diffusion between ions of the same charge is significantly lower than their individual diffusion, with the most notable difference observed for LIL$^-$LIL$^-$ diffusion. Apparently, this difference arises from the negative contribution of correlated ion diffusion, where cation motion dominates anion motion by approximately 4-20%. Specifically, the absolute difference between correlated and individual cation diffusion ranges from 20% (LIL4) to 40% (LIL2 and LIL6), whereas for anions, this difference does not exceed 20% for LIL2 and LIL6 and is almost negligible for LIL4. The remaining diffusion term, formed by the correlated motion of oppositely charged ions, predominates over individual diffusion by an order of magnitude and approximately by two orders of magnitude over collective diffusion.

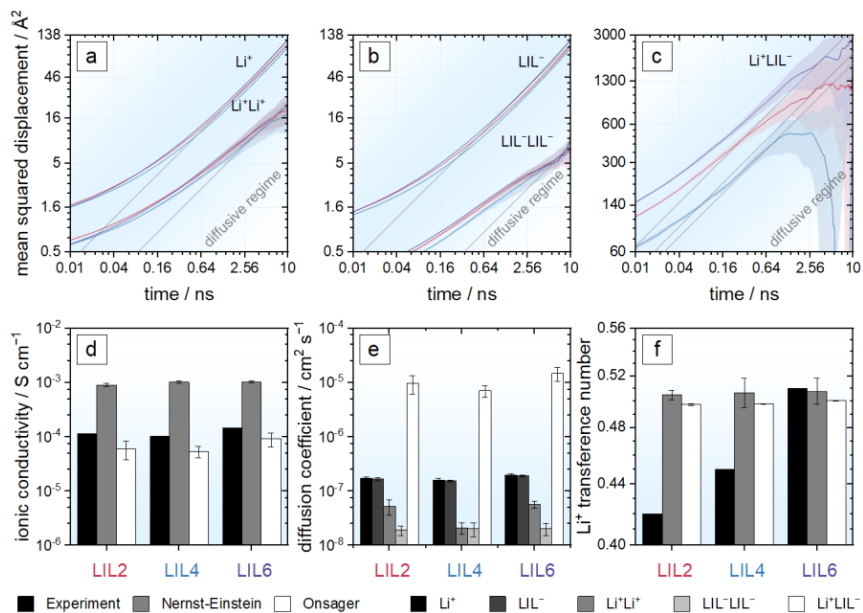

**Figure 1.** Illustration of ion mobility characteristics, represented by an average (**a-c**) mean squared displacement describing individual and collective ion motion of (**a**) Li$^+$, (**b**) LIL$^-$ and (**c**) Li$^+$ coupled with LIL$^-$ in LIL2, LIL4 and LIL6 obtained by means of averaging over ten independent runs. Color contours are the standard deviation from the average; the gray lines are drawn at a 45° for visual guidance and represent the slope at which diffusion regime occurs; (**d, f**) ionic conductivity and transference number against experimental measurements,[19] and (**e**) individual and collective diffusion coefficients of the cation and respective LIL anion calculated using Nernst-Einstein equation and its expression enhanced by Onsager transport coefficients.[33] Error bars represent standard deviation from the average.

Overall, this observation is well corroborated by above given conclusion on the role of correlated ion contribution to the conductivity. Individual ion mobilities predominantly determine the collective mobilities of ions of the same charge, with correlated motion playing a smaller yet positive role in ionic conductivity. This balance, crucial for positive and negative ion mobilities, is expressed experimentally through transference numbers, indicating the fraction of electric current carried by a specific cation and serving as a measure of electrolyte efficiency for LIBs. Despite theoretical and experimental differences, particularly for LIL2 and LIL4, Figure 1f demonstrates the significant role of individual ion mobilities in terms of their collective behavior, with transference number differences between Nernst-Einstein and Onsager approaches not exceeding *ca* 2%. Correlated diffusion of counterions, due to opposite charges, negatively impacts total ionic conductivity, having the most notable effect compared to other terms. This higher impact may stem from the absence of ion association typical in conventional electrolytes. On the other hand, it may be assumed that saturation of LIL with fluorinated groups reduces ionic conductivity due to variations in $Li^+$ coordination ability and resultant inhomogeneous cation mobilities across different regions of the anion.

Indeed, inhomogeneity (or heterogeneity) in ion dynamics, expressed by the second-order non-Gaussian parameter, can provide a quantitative assessment of the impact of different spatial regions of LILs on the above discussed dynamic characteristics.[35] In our case, asymmetric distribution is observed across all LILs, with the maximum height around 0.2 at 0.1 ns (Figure 2a-c). More precisely, the relatively small $Li^+$ probability comparing to one in other common electrolytes ranks as follows: LIL4 (0.2099 at 0.11 ns) > LIL2 (0.2074 at 0.09) > LIL6 (0.2053 at 0.08), indicating an increasing deviation from a homogeneous behavior. On the other hand, the position of the peak indicates the time of onset of diffusive dynamics after the ballistic stage. Interestingly, despite structural similarities between LIL2 and LIL4, differing from LIL6, this behavior suggests that a distinct ion transport mechanisms arise due to varied electron withdrawing groups. Notably, $Li^+$ shows higher dynamic heterogeneity compared to $LIL^-$, suggesting a higher sensitivity of the cation to local environment.

Another level of insight into the dynamic sensitivity to the surrounding environment can be gained by analyzing self-part of van Hove correlation function, which represents the time dependent individual ion displacement with respect to its initial position.[36] Figure 2d-f shows that, across all considered LILs, $Li^+$ displacement probabilities generally follow similar trends over time. For short intervals, probabilities are more skewed compared to longer simulation times. Notably, in LIL4 $Li^+$ demonstrates longer displacement times (~20-30 ns) compared to LIL2 (~20-25) and LIL6 (~20-21), respectively. A more careful examination of distribution shape reveals subtle deviations, which are more pronounced for LIL4 and LIL2 (see FigureS2-3 for corresponding profiles).

Focusing on a single arbitrary cation displacement, rather than averages, may shed a light on such uncertainty.[16] Thus, Figure 2g-j illustrates a discontinuous behavior of the $Li^+$ distribution in LIL2 with jumps occurring at ~20-30 ns over the distances of 8-12 Å, while LIL4 covers wider distances and jumps at ~30-35 ns for *ca* 8 Å. In contrast, the distribution of cation in LIL6 is more continuous along the entire trajectory and $Li^+$ displacement not exceeding *ca* 16 Å. Based on this, it can be assumed that during the pathway of the cation it tends to change the local environment/coordination structure in a favor of diffusive motion, particularly in LIL2 and LIL4. In this regard, to get a comprehensive pattern of $Li^+$ dynamic displacement one needs to consider their collective motion along the trajectory path.

This task involves the analysis of distinct parts of the van Hove correlation function, which, unlike the above analysis, can access collective, *pair*, ion displacement, providing information about how distant certain ions are from each other at given time intervals. As illustrated in Figures 2k-m, the resulting $Li^+Li^+$ displacement shows two distributions: the first one is located at lower distances, with its extension not exceeding 7-8 Å, while for the second, less prominent, the extension of the distribution barely covers 5 Å. Throughout the simulation time, the intensity of the former distribution increases, and its maximum shifts towards lower distances, which is more pronounced for LIL4 and LIL2. In contrast, the latter distributions, *i.e.*, those located at 9.29, 9.38, and 10.07 Å in LIL2, LIL4, and LIL6, shift to larger separation distances and decay at later time intervals (see also FigureS4 for details). This indicates that along the simulation time, the distance between $Li^+Li^+$ pairs is decreasing, weakening pair dynamics, which is prominent in LIL2 and LIL4. On the other hand, the increasing of another peak indicates that the position of the first ion is occupied with high probability by another one.

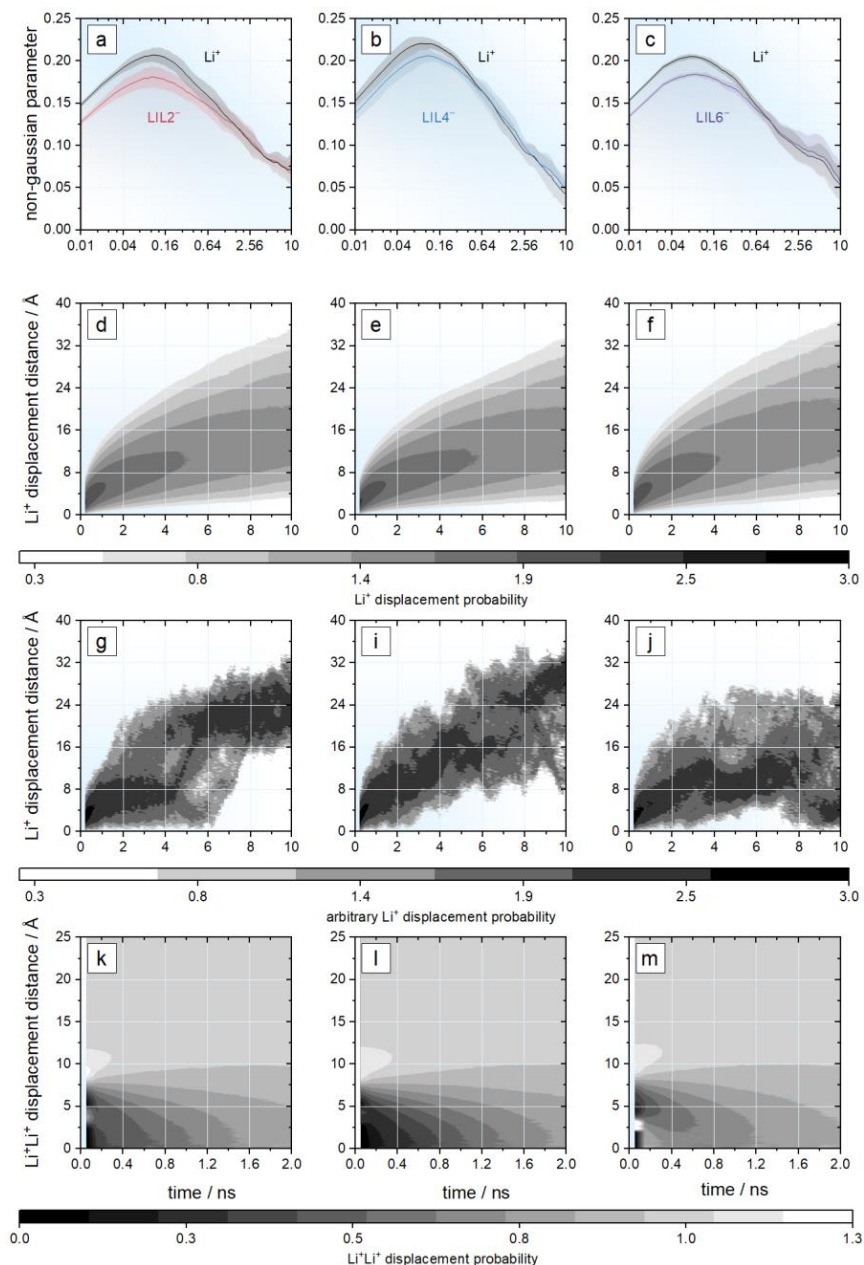

**Figure 2.** Illustration of inhomogeneity in ion mobility, represented by an average (**a-c**) non-Gaussian parameter for $Li^+$ and $LIL^-$ in LIL2 (*left*), LIL4 (*middle*) and LIL6 (*right*) obtained by means of averaging over ten independent runs. Color contours in are the standard deviation from the average; probability of (**d-f**) average individual and (**g-j**) single arbitrary $Li^+$ displacement, as well as (**k-m**) average collective $Li^+$ displacement expressed by self- and distinct part of van Hove correlation function.[36]

Thus, assuming the key role of the local environment in ion dynamics, the above observations can be rationally explained. The difference in electron-withdrawing groups (read electronic structure) of LILs suggests the difference in coordination affinity of the cation and anion, and hence, the inhomogeneity of their dynamics. Particularly, for the cation, this may be reflected in the formation of the coordination cell around it, which hinders its diffusive motion. Furthermore, an opposite to $Li^+$ diffusion behavior of the non-Gaussian parameter for the respective LILs, *i.e.*, LIL4 > LIL2 > LIL6, also suggests slow diffusive motion of the coordination cell, which is more pronounced for LIL4 with lower conductivity. On the other hand, along the coordination cell dynamics, the cation tends to escape from this surrounding environment. However, for LIL6, such tendencies are hardly noticeable, and $Li^+$ transport follows vehicular mechanism, whereas in LIL2 and LIL4, the cations are more caged, leading to random jumps along the ion pathway. The latter is confirmed by the analysis of the pair $Li^+Li^+$ displacements, where upon the displacement of one cation, the second one tends to jump to the vacant position previously occupied by the first particle as indicated by the valley arising near the first peak at larger time intervals. In this regard, the LIL4 cation takes a longer time to occupy the free vacancy compared to LIL2, which reaches the vacant position faster. Apart from the fact that the individual displacement of an arbitrary cation in LIL6 does not show any prominent evidence of such jumps, from the collective point of view, it also tends to exhibit jump-like motion along the vehicular pathway.

Given the ionic nature of LILs and taking into account the above suggestion that the ionic conductivity determined by the $Li^+$ dynamics across the different local regions of the anion, the analysis of ion domains is of particular interest. Indeed, upon visually assessing representative MD cell configurations (Figure 3a-c), it becomes apparent that it is segregated into several spatial regions among which one can clearly distinguish areas reached by ethoxide and fluorinated groups, conditionally different in polarity. Moreover, for LIL2 and LIL4, such spatial regions are more pronounced due to the larger number of trifluoromethyl groups, which form channels separating ethoxide regions.

Statistically, these visual features can be expressed through domain formation distribution, describing the probability of the mutual occurrence of ions of different types within the same aggregate (Figure 3d-f). Similar to other systems,[37, 38] the resulting formation mainly consist of charged isolated $Li^+$ and $LIL^-$, as well as neutral ion pairs. Along with this, there is also a probability of forming larger domains, which decreases with increasing of their spatial extension. This is particularly noticeable in the cases of LIL2 and LIL4, which compose more than 8 cation and anions. Despite the negligible probability of such aggregate formation, they promote an imbalance between ions of different charges. In this context, LIL4 shows a higher deflection from neutrality towards positively charged aggregates (23.58%) compared to LIL2 (14.43%) and LIL6 (9.47%), implying the presence of Li-rich domains in the system. This observation is also supported by the probability of the existence of isolated ions and neutral pairs. In both cases, LIL4 takes a leading position over other LILs, demonstrating a higher number of neutral pairs and a higher order of imbalance in favor of isolated anions, compared to other LILs. Considering different sites of the anion separately, we can observe a generally similar tendency of domain asymmetry, higher for LIL4 (20.90%) and lower for LIL2 (13.11%) and LIL6 (11.14%), when only oxygen atoms are selected, and a somewhat non-trivial pattern for fluorine atoms, where lower asymmetry is inherent for LIL2. Apart from these differences, it is worth noting the difference in asymmetry with respect to the different sites of the anion, where the positive region between the domains involving oxygen and fluorine atoms of the anion is much more pronounced for LIL6. Based on this observation, it can be assumed that such tendencies may be associated with the different spatial extents of ion domains capable of containing a different number of cations.

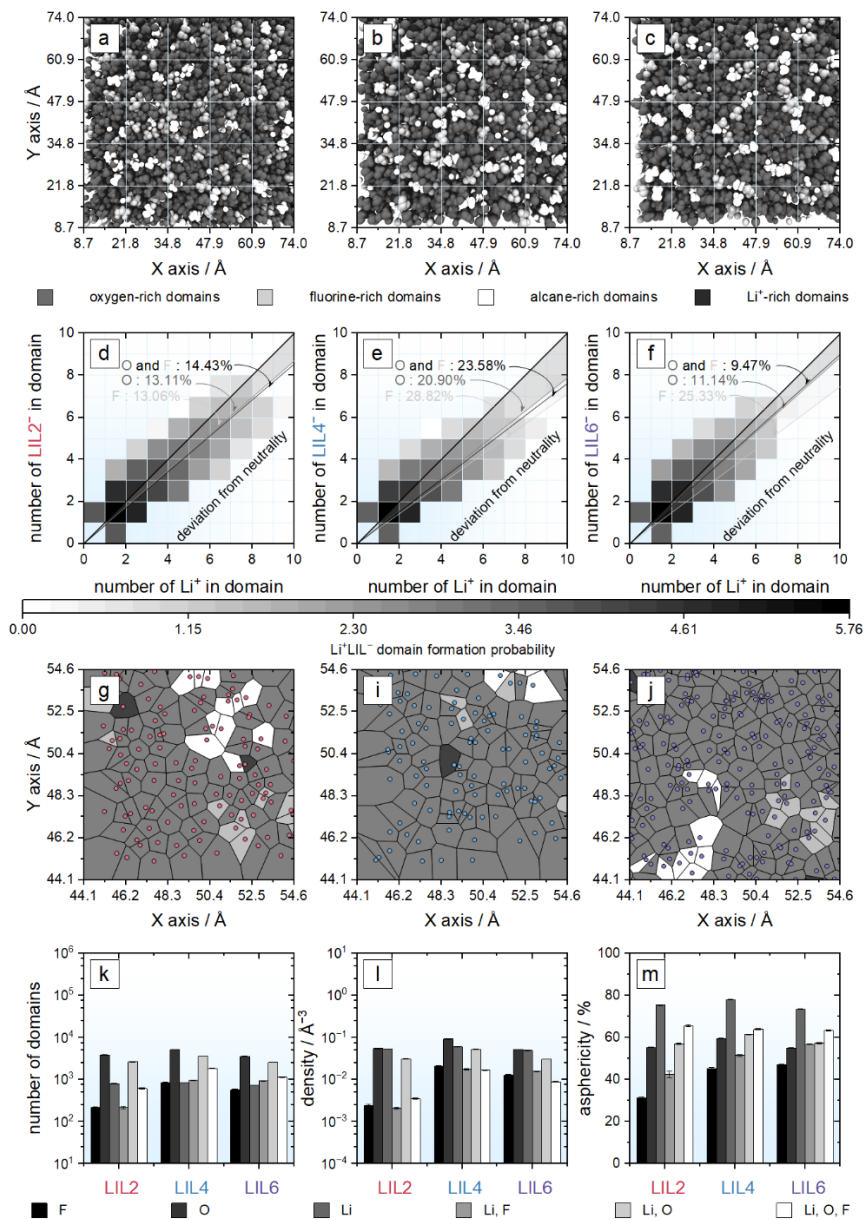

**Figure 3.** Illustration of inhomogeneity in local structure, represented by an average (**a-c**) snapshots LIL2 (*left*), LIL4 (*middle*) and LIL6 (*right*) obtained after the equilibration run; (**d-f**) domain formation probability of LIL$^-$ with respect to Li$^+$. The horizontal bar represents the occurrence frequency of the corresponding domains analyzed based on threshold distance of 3.75 Å. Mind the logarithmic scale. The grey diagonal line represent neutrality of the domain; lines represent deviation from neutrality toward positive ion rich domains (in %) using different criterion of domain formation[37] (see Figure S5 for details); (**g-j**) Voronoi tessellation, describing the spatial organization of polyhedra representing different domains. Dots represent the atomic centers; (**k-m**) Voronoi metrics represented by the (**k**) number of domains, (**l**) spatial density of the polyhedron and (**m**) asphericity parameter. Error bars represent standard deviation from the average.

The latter may be quantitatively estimated within a framework of Voronoi tessellation analysis.[39, 40] In general, this analysis implies description of the local structure by means of a spatial partitioning into regions, *domain*, based on proximity to a set of points, hence dividing space into tessellation (see Figure 3g-j) that can be described statistically. Among the variety of statistical metrics used previously for this purpose,[41, 42] here we considered the average number of domains that counts the regions at each point, density, which measures the space available per region, and the asphericity, which assesses how compact or irregular these regions are relative to their boundaries. Therefore, following the statistical pattern in Figure 3k-m, all metrics generally exhibit the trend LIL4 > LIL2 > LIL6 across domain types. However, some inconsistencies emerge, notably with fluorine-rich domains deviating from the trend, where LIL4 remains dominant, while, similar to above analysis, LIL6 found to be higher than LIL2. Indeed, considering all atoms as domain formations reveals a local structure in LILs with relatively compact and highly skewed cells, particularly evident for LIL4. This trend persists when lithium interacts solely with fluorine, showing lower cell asphericity indicating a more ordered yet diverse arrangement. Oxygen, when coupled with lithium, notably shifts the balance towards LIL2 while maintaining LIL4 dominance, distorting local organization and making cells less skewed and packed, thus rendering LILs more heterogeneous and less uniform. Oxygen-rich single-atom domains are widespread and tend to occupy dispersed regions, contrasting with slightly more compact and less skewed Li-rich regions, which have less space per domain. This trend is most pronounced for LIL4 compared to slightly smaller LIL2 and LIL6. Finally, fluorine single-atom domains are strongly skewed shapes and less dispersed within the LILs compared to other regions.

From the above analysis, it can be suggested that the higher extension of spatial regions, *i.e.*, inhomogeneity, is inherent in LILs containing more trifluoromethyl groups. Fluorine-rich domains stand as stabilizers that exhibit less dispersion compared to other regions and, we assume, do not participate in interaction with $Li^+$ ions, thus having a low probability of being within its coordination shell. As a result of such non-interference, different LILs introduce an imbalance toward positively charged domains, thereby changing the overall dominant trend between LIL2 and LIL6 (while LIL4 maintains a leading role). Along with $Li^+$, oxygen-rich regions also influence the overall structural heterogeneity, making them more dispersed and compact. In this case, the general tendency for both types of domains is similar to that observed in the dynamic heterogeneity analysis, *i.e.*, LIL4 > LIL2 > LIL6, suggesting the deceleration of ion dynamics and the presence of hindrances along the transport pathway.

Given the key role ascribed to structural heterogeneity in governing ion transport, the above domain analysis warrant examination in terms of the radial and coordination features of the underlying interactions. Since we assume that ion dynamics is determined by the cation, we, therefore, consider here Li as a reference site. As it might be seen from Figure 4a, resulting curves, reflecting the pair distances probability between Li and O, exhibit a well-defined first peak at *ca* 1.35 Å with a broad valley defining the border of solvation shell. A more careful examination of the peak position shows a slightly lower distances for LIL6 compared to LIL2 and LIL4. Further examination of the related distances but with different type of oxygen atoms, *i.e.*, covalently bonded to boron atoms, and ones belongs to ethoxide group, revealed the noticeable difference in the radial probabilities and the respective tendencies of increasing/decreasing. Nevertheless, due to similarities in peak positions it is hard to draw any conclusions regarding the preferential localization of the cation.

Therefore, the strength of the interactions between the atomic sites was probed by the average distance between the reference and the first neighboring site.[43] In this regard, it was found that despite the relatively high positive charge of boron atom, cation demonstrates a preferential localization around covalently attached oxygen atoms, particularly for LIL6 (1.61 Å) compared to LIL4 (1.88) and LIL2 (1.75). For the ethoxide oxygen the strength of the interactions was found to be weaker 2.23-2.47, whereas its tendency for the different LILs follows the reverse order from LIL2 to LIL6. The variation in strength of interactions between cation and different oxygen sites is similarly reflected in the respective coordination numbers, when integrating the radial distribution within the border of the solvation shell (Figure S6). Such trend is also well corroborated with the opposite behavior of radial distributions between the cations, where a clearly distinct peak is observed for LIL6 (Figure 4b).

Considering the contribution of fluorine groups into the Li$^+$ coordination environment, it was found a quite different behavior of radial distributions. From the point of general tendencies, it is clearly seen in Figure 4c that the probabilities of these distributions are quite low compared to distributions observed above, which may be interpreted as a quite poor contribution of fluorine into the Li$^+$ solvation shell. This also confirmed by the respective average distances, where among all the systems LIL6 found to be the lowest one – 4.36 Å, compared to LIL2 (3.86) and LIL4 (4.14).

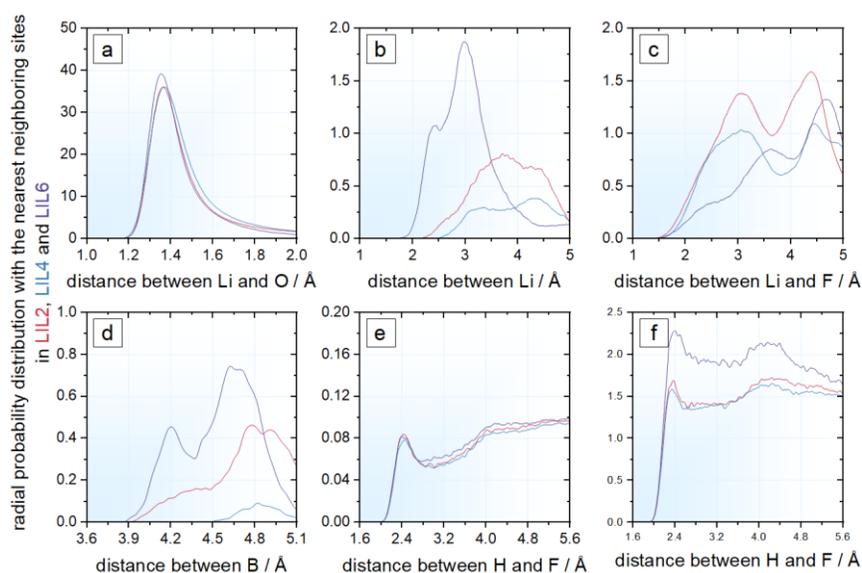

**Figure 4**. Illustration of the local structure organization, represented by the nearest neighboring radial distribution describing the strength of the interactions between (**a-c**) Li$^+$ and corresponding atoms of LIL$^−$, and (**d-f**) between the atoms within different LILs.

Assuming the self-association of the anions, the respective radial distributions were also analyzed to reveal the impact of fluorine groups (Figure 4d-f). From the generalized point of view, the most probable interactions occur for the LIL6 and LIL2 compared to LIL4. In addition, it was also found the absence of any localized hydrogen bonding interactions between the anions, while the appearance of the peak at 2.4 Å, assumed to be a result of steric effects (Figure 4e-f). For the pairs involving fluorine atoms a slightly intensive double peaks at 2.4 and 4.2 Å also does shows any prominent interactions, while such interactions are slightly stronger compared to ones with oxygen site. On the other hand, a slightly higher intensities for LIL6 may explain the weak participation of fluorine atoms in Li$^+$ coordination shell.

In overall, it is confirmed that the variation of electron-withdrawing groups in three LILs differently alters their local structural organization. Upon saturation of electron-withdrawing groups by fluorine atoms from LIL6 to LIL2 and LIL4, the coordination structure becomes less significant, leading to weakened interactions between ions of the same charge. Moreover, a higher number of fluorine atoms in the LIL structure is reflected in higher coordination of Li$^+$ along the ethoxide chains and lower coordination in proximity to boron atoms. This structural pattern segregates the LIL into different regions, with an essential predominance of the regions containing the cations coordinated by oxygen atoms. The higher dispersion and density of these regions, as well as their higher positive charge, lead to hindrances along the cation pathway, thus making it more intermittent. Slowing the ion motion, as a result of these transport features, governs the extent of correlated motion between ions of the same charge, as well as between the counterions, significantly impacting total ionic conductivity. Thus, understanding how the variation of electronic structure affects the local structural organization of LILs provides crucial insights for designing electrolytes with enhanced ionic conductivity and improved performance for LIBs.

**Acknowledgments**

This work was supported by the by the Assistant Secretary for Energy Efficiency and Renewable Energy, Office of Vehicle Technologies of the US Department of Energy, through the Battery Materials Research (BMR) program. We gratefully acknowledge the computing resources provided on Bebop, the high-performance computing cluster, operated by the Laboratory Computing Resource Center at Argonne National Laboratory.

Supporting Information

**Impact of Electron-Withdrawing Groups on Ion Transport and Structure in Lithium Borate Ionic Liquids**

Volodymyr Koverga, Selvaraj Selva Chandrasekaran and Anh T. Ngo[*]

Department of Chemical Engineering, University of Illinois Chicago, Chicago, IL 60608, United States
Materials Science Division, Argonne National Laboratory, Lemont, IL 60439, United States

[*]Corresponding author's email: anhngo@uic.edu

For classical MD simulation, three LILs composing anions based on hexafluoro-2-propanoxy (LIL2), hexafluoro-2-methyl-2-propanoxy(LIL4) and trifluoro-2-propanoxy (LIL6) derivatives were selected. The non-polarizable all-atomic OPLS(-AA), "Optimized Potentials for Liquid Simulations",[1, 2] force field was utilized in the framework of Moltemplate code[3] to adjust most of the intra- and interatomic potential parameters, describing covalent terms within bond stretching, angle bending and dihedral angle torsion along the covalent bond, and non-covalent terms represented by the van der Waals interactions. The Lennard-Jones parameters for B−O were adopted from Wang et al.[4] and Li was taken from the work of Farhadian and Malek[5]. The Coulomb interactions were described by the partial charges evaluated in the framework of CHELPG, "CHarges from ELectrostatic Potentials using a Grid-based method",[6] molecular electrostatic potential scheme using MP2/6-31++G(d, p) level of theory for the set of LIL's ion pairs geometries, preliminary optimized in HF/6-31G(d) as available in Gaussian code, v16.[7] In addition, each of the isolated LIL anion was optimized to determine electrostatic potential using the same approach. To boost the ion dynamics the Lennard-Jones parameters were modified, particularly, $\varepsilon$ parameters were reduced for Li and H atoms ($\varepsilon_{Li}$ = 0.0694 and $\varepsilon_H$ = 0.0061 kcal mol$^{-1}$) to reduce the energetics in the interactions between the counterions and between the ethoxide group of the anion. The comparison of Li-ion dynamics using different electrostatic potentials is illustrated in Figure S1 by time-evolution of mean squared displacement.

Optimized geometries were additionally checked to be in true minima by the absence of imaginary frequencies in the corresponding vibration spectra. Missing in database intraatomic potentials (except B−O and O−B−O that was taken from literature[4]) for atoms covalently bonded to central B atom of the anion were reparametrized using relaxed potential energy scan concept, $r$PES. To speed up the calculations all the LIL geometries were reduced by two ethylene oxide units from each side and reoptimized in the gas phase using the same level of theory as for the full geometries. Next, a series of 10 scans was conducted to evaluate stretching and bending parameters using a step of 0.04 Å and 0.4°, respectively. For torsion potentials 18 scans with a step of 10° were implemented. The obtained in this way $r$PES profiles were averaged over all the degrees of freedom of the same nature and subsequently fitted by suitable polynomials according to OPLS-AA analytical expression of the potential energy adopted for LAMMPS environment.

$$U = \sum_{bond} k_r \Delta r^2 + \sum_{angle} k_\theta \Delta \theta^2 + \sum_{dihedral} \sum_{n=0}^{5} A_n \cos^{n-1}(\phi)$$

where $U$ is a sum over the internal terms as a function of atomic coordinates represented by bond distances ($r$), angles ($\theta$), and dihedrals ($\phi$); The parameters $k_n$ and $A_n$ are the respective force constants and the variables.

The initial coordinates of 864 ion pairs were generated using Packmol code[8] and placed into orthorhombic supercell of $100 \times 100 \times 100$ Å with three-dimensional periodic boundary conditions. Such system size was selected to avoid deleterious pressure fluctuations and to reduce the influence of the on finite size effects on the electrostatic interactions during the equilibration stage, while a considerable larger box was taken to avoid the intermolecular clashes. All MD simulations were carried out using LAMMPS, v080223.[9] Equations of motion were integrated with a time-step of 2 fs. The time-step selection was justified by utilization of C−H bonds constrains, which was achieved with SHAKE, "Spherical Harmonic Accelerated Kinetic Energy", algorithm.[10] The electrostatic long-range interactions within the cut-off range of 12 Å were accounted by the computationally efficient Particle-particle-particle-mesh method was used to evaluate electrostatic energies (the accuracy of $10^{-5}$), using the same cut-off distance for the real-space component.

The simulation protocol implied a stepwise equilibration procedure, followed by sampling the coordinates for the analysis. Thus, each system was, firstly, minimized using steepest descent algorithm with the default convergence criterion, next, a series of equilibration steps in an isothermal-isobaric (*npT*) and canonical (*nVT*) ensembles was conducted to reach experimental density of ~1.021±0.006 g cm$^{-3}$ [11] and relax the LIL geometry:

1. 0.1 ns of *npT* compression at 100 atm and 373 K using Nose-Hoover thermostat and barostat[12-14] with coupling constants, $\tau$, of 300 fs and 800 fs;
2. 0.5 ns of *npT* compression at 100 atm and 373 K;
3. 0.5 ns of *npT* compression at 100 atm and 373 K;
4. 1 ns *nVT* heating at 603 K;
5. 1 ns of *npT* compression at 100 atm and 298 K;
6. 5 ns of *npT* compression at 1 atm and 298 K to estimate equilibrium density;
7. 10 ns of *nVT* production using Nose-Hoover thermostat to collect the coordinates for the analysis of the local structure organization;
8. 50 ns of *nVT* to collect the coordinates for the analysis of ion dynamics

Obtained trajectory(-*ies*) at Step 8 was used to estimate ionic conductivity with Nernst-Einstein equation:

$$\sigma = \frac{e^2}{6k_B VT}(NqD_{Li}^{self} + NqD_{LIL}^{self})$$

and using Onsager transport coefficients:

$$\sigma = (\sigma_{Li}^{self} + \sigma_{Li}^{distinct}) + (\sigma_{LIL}^{self} + \sigma_{LIL}^{distinct}) - 2\sigma_{Li,LIL}$$

$$\sigma^{self} = \lim_{x \to \infty} \frac{e^2}{6tVk_BT} \sum_{i=1}^{N} <|\mathbf{r}_i(t)|^2>$$

$$\sigma^{distinct} = \lim_{x \to \infty} \frac{e^2}{6tVk_BT} \sum_{i \neq j}^{N} <\mathbf{r}_i(t)\mathbf{r}_j(t)>$$

where $\sigma^{self}$ and $\sigma^{distinct}$ are individual, *self*, and collective, *distinct*, contributions of Li and LIL ions to ionic conductivity – their sum represents the total collective contribution of Li and LIL respectively, $e$ is the elementary charge, $k_B$ is the Boltzmann constant, $V$ is the system volume, $T$ is the temperature, $N$ is the number of charged species of the same, $N_i$, or different, $N_iN_j$ ($i \neq j$), sort, $<\mathbf{r}(t)>$ is an ensemble-averaged difference between vector position of species $i$ or $i$ and $j$ at time $t$, $D$ is self or distinct diffusion coefficient and $q$ is total charge of Li or LIL ions. The ionic conductivity or self-diffusion and were estimated based of Fickian formalism, according to which $<\mathbf{r}(t)> \propto t^{\beta}$, where $\beta$ was estimated using the equations:

$$\beta^{self} = \frac{d\log<|\mathbf{r}_i(t)|^2>}{d\log t} = 1$$

$$\beta^{distinct} = \frac{d\log<\mathbf{r}_i(t)\mathbf{r}_j(t)>}{d\log t} = 1$$

where $\beta = 1$ represent the slope of 45° where diffusion regime occurs. Obtained transport characteristics were used to calculate Li transference number using both Nernst-Einstein and Onsager approximations:

$$t_{Li} = \frac{qNF^2}{RT}(D_{Li}^{self} + D_{LIL}^{self})$$

$$t_{Li} = \frac{\sigma_{Li}^{self} - \sigma_{LIL}^{self}}{\sigma_{Li}^{total} + \sigma_{LIL}^{total} - 2\sigma_{Li,LIL}}$$

Dynamic inhomogeneity (heterogeneity) was analyzed using second-order non-Gaussian parameter, $\alpha_2$:

$$\alpha_2(t) = \frac{3<|\mathbf{r}(t)|^4>}{5<|\mathbf{r}(t)|^2>^2} - 1$$

In addition, van Hove correlation function was used to describe the correlation between two species in time and space:

$$G(r,t) = \frac{1}{4\pi\rho Nr^2}\sum_{i,j}\delta(r-|\mathbf{r}_i(0)-\mathbf{r}_j(t)|)$$

where $\rho$ is atom number density, $r$ is the distance. Next, van Hove function was decomposed by two parts – *self*:

$$G^{self}(r,t) = \frac{1}{4\pi\rho Nr^2}\sum_{i}\delta(r-|\mathbf{r}_i(t)|)$$

and *distinct*:

$$G^{distinct}(r,t) = \frac{1}{4\pi\rho Nr^2}\sum_{i\neq j}\delta(r-|\mathbf{r}_i(0)-\mathbf{r}_j(t)|)$$

Local structure analysis, particularly radial distributions, discussed in this work was performed using TRAVIS, "TRajectory Analyzer and VISualizer" code, v062922[15, 16].

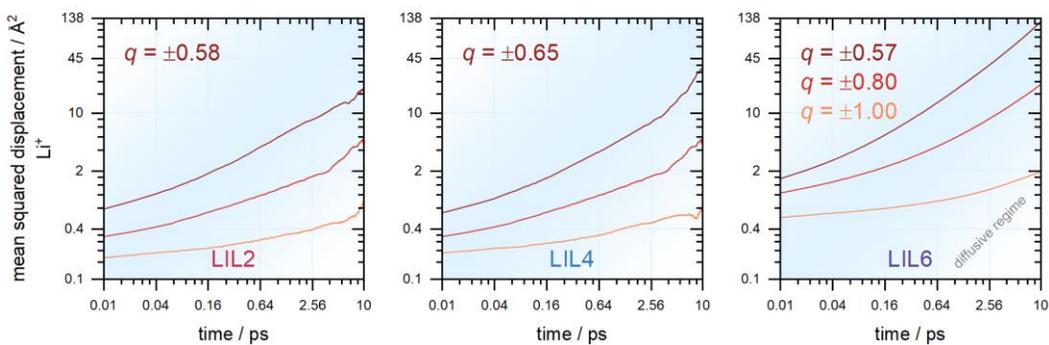

**Figure S1.** Illustration of mean squared displacement describing individual, *self*, Li$^+$ motion in LIL2, LIL4 and LIL6 calculated using three charge distribution approximations: isolated ion pair, isolated anion and scaled by a factor 0.8.

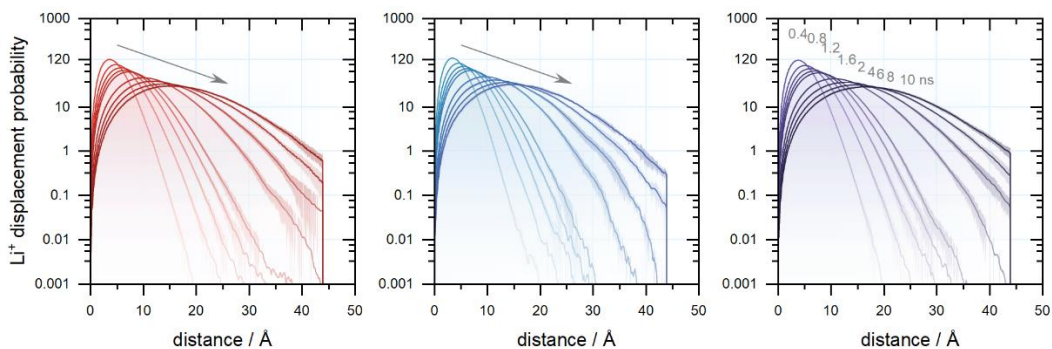

**Figure S2.** Illustration of the evolution of Li$^+$ displacement probability expressed by the self-part of the van Hove correlation function on selected time intervals for LIL2, LIL4, and LIL6.

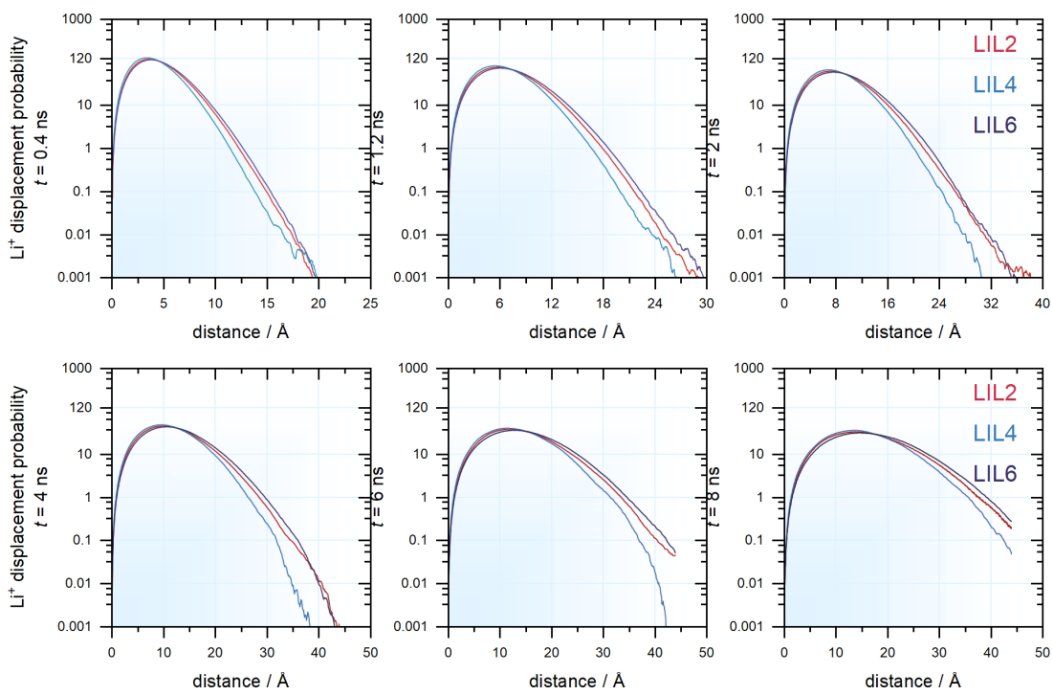

**Figure S3.** Illustration of the evolution of Li+ displacement probability expressed by comparing the self-part of the van Hove correlation function between LIL2, LIL4, and LIL6 at selected time intervals.

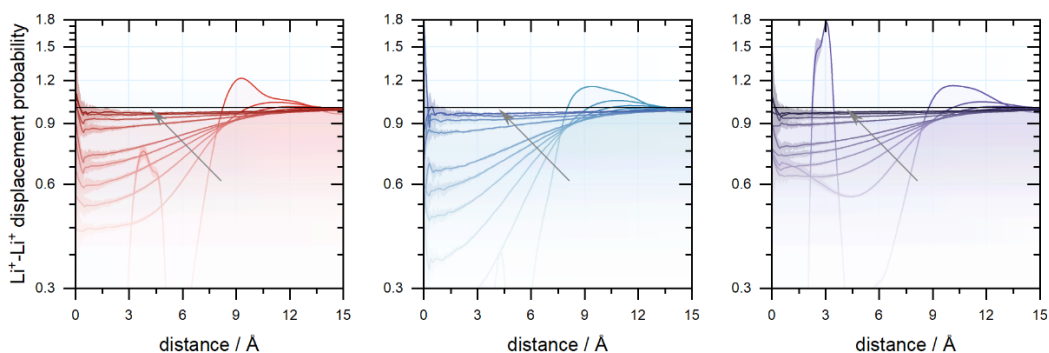

**Figure S4.** Illustration of the evolution of Li$^+$Li$^+$ displacement probability expressed by the distinct-part of the van Hove correlation function on selected time intervals for LIL2, LIL4, and LIL6.

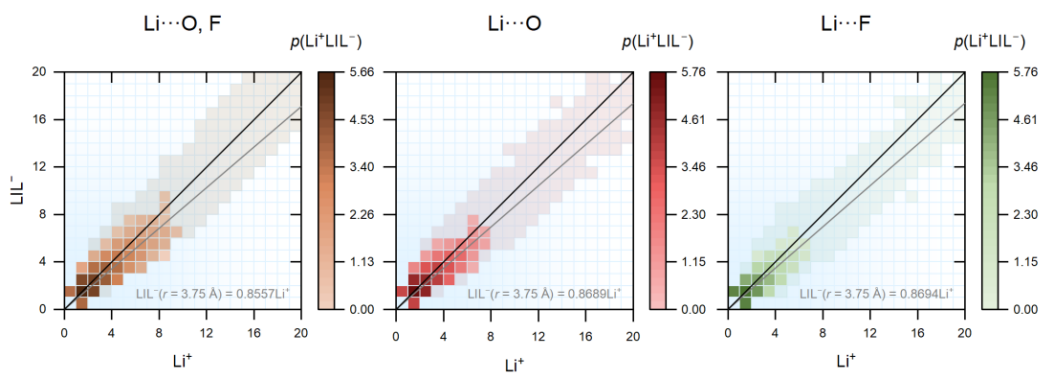

**Figure S5.** Illustration of domain formation probability of Li$^+$ with respect to LIL2$^-$. The black diagonal line represents neutrality of the domain, and the grey line represents deviation from neutrality toward positive ion-rich domains fitted to the first-degree polynomial LIL$^-$ = $a$Li$^+$ (where $a$ is a fitting coefficient) using different criteria of domain formation between Li atoms and O, F, and both O, F atoms. Larger distribution profiles illustrate an example of domain distribution using a threshold distance of 5 Å between atomic sites.

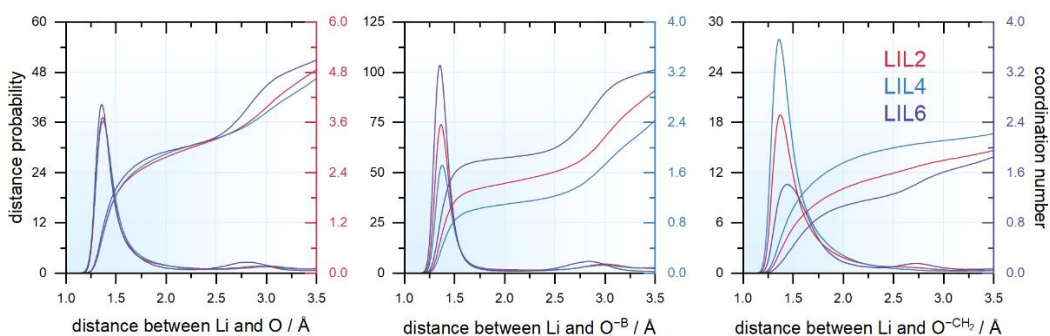

**Figure S6.** Illustration of the radial distribution and running coordination number, describing the strength of the interactions between Li$^+$ and generalized representation of oxygen atoms, as well as different type of oxygen atoms of LILs, and Li$^+$ local coordination environment.